\documentclass[%
 reprint,
 amsmath,amssymb,
 aps,
]{revtex4-2}

\usepackage{graphicx}
\usepackage[utf8]{inputenc}
\usepackage{dcolumn}
\usepackage{bm}

\usepackage{xcolor} 

\begin{document}

\preprint{APS/123-QED}

\title{Stochastic Cluster Expansion for Excited State Energies}

\author{Annabelle Canestraight}
 \affiliation{Department of Chemical Engineering, University of California, Santa Barbara, CA 93106-9510, U.S.A.}
\email{acanestraight@ucsb.edu}
\author{Russell Miller}
\affiliation{Department of Chemistry and Biochemistry, University of California, Santa Barbara, CA 93106-9510, U.S.A.}
\author{Libor Veis}
\affiliation{Department of Theoretical Chemistry, J. Heyrovský Institute of Physical Chemistry, Czech Academy of Sciences, Prague 18223, Czech Republic}
\author{Vojtech Vlcek}
\affiliation{Materials Department, University of California, Santa Barbara, CA 93106-9510, U.S.A.}
\affiliation{Department of Chemistry and Biochemistry, University of California, Santa Barbara, CA 93106-9510, U.S.A.}

\date{\today}

\begin{abstract}
Excited-state electronic structure in strongly correlated systems remains challenging due to the exponential scaling of the many-body Hilbert space and the difficulty of constructing systematically controlled active spaces. Building on the stochastic cluster expansion (SCE) framework previously developed for ground-state correlation energies, we extend the formalism to excitation gaps by expressing energy differences directly as a hierarchy of orbital-space cluster contributions. In this formulation, excitation energies are reconstructed from reduced-rank calculations involving a minimal frontier chemical subspace (FCS), treated exactly, together with stochastic sampling of the remaining orbital environment. This approach eliminates the need for large or chemically preselected active spaces. We demonstrate the method on charge-transfer complexes and polyacenes, where accurate singlet-triplet gaps are obtained that agree with full-system results. The method converges with low-order cluster terms and provides a systematically improvable framework for excited states in correlated systems.
\end{abstract}

\maketitle

Excited-state electronic structure plays a central role in a wide range of chemical and physical processes, including fluorescence in organic fluorophores~\cite{kasha1950characterization, kasha1960paths, Lakowicz2006}, the optical response of biological chromophores~\cite{Ma2010Modeling}, and the development of high-efficiency optoelectronics~\cite{Zimmerman2010, Uoyama2012}, with many key applications hinging on singlet–triplet energy gaps. Accurate determination of excitation energies in correlated systems therefore remains a fundamental challenge in electronic structure theory. In principle, these quantities can be obtained directly from exact diagonalization (ED), which provides both eigenvalues and eigenvectors of the many-body Hamiltonian. In practice, however, the exponential growth of the Hilbert space renders ED intractable even for relatively small conjugated systems such as polyacenes, which have become the de-facto system for assessing correlated methods\cite{DavidSherrill1999,eriksen2020shape,ST-delicate,ST-diradical,ST-reichman,C8SC03569E,ghosh2017generalized}.

To overcome this limitation, considerable effort has been devoted to the construction of reduced active spaces in which high-level many-body solvers can be feasibly applied\cite{NormSquishActiveSapce,WeirdAlgorithmiq,He2020-raFOEmbedding,Muchler2022Static,GwenReducedScaling,GwenEfficientTreatment,Bartlett2007,reiher_perspective,DMRG_White,OlivaresAmaya2015,sayfutyarova2017automated,bao2019automatic}. This approach relies on the assumption that correlation is predominantly localized within a certain subset of single-particle orbitals. In such a reduced active space, electrons are allowed to undergo excitations, while those outside are treatable by mean-field methods. This approximation is well justified when the dominant correlation effects are associated with a limited set of near-degenerate or partially occupied orbitals, as is often the case in strongly correlated or chemically active subspaces. More generally, the notion of ``correlation localization'' is basis dependent and not necessarily spatial; for example, correlation may be concentrated within an energy window \cite{SakumaWerner2013Dynamical,Aryasetiawan2009,MiyakeEffevtiveBand,Romanova2023,GwenReducedScaling,Muchler2022Static}. For instance, in conjugated molecules, low-lying excitations are often dominated by the $\pi$ orbital manifold, with $\sigma$ orbitals contributing primarily as a mean-field background. This separation has enabled accurate treatment of ground and excited states using methods such as density matrix renormalization group (DMRG), coupled cluster, auxiliary field quantum Monte Carlo (AFQMC), and related approaches~\cite{Szalay2015,legeza_2003a,Legeza2003,MOLMPS,chan_review,ST-diradical,ST-reichman,AFQMCEskridge2019,shee2024staticquantumembeddingscheme}. In more complex systems, where the relevant active space cannot be identified from symmetry or simple chemical intuition, approaches such as CASSCF or orbital localization techniques are employed to construct reduced spaces~\cite{WeirdAlgorithmiq,malmqvist1989casscf,helmich2019benchmarks,olsen2011casscf}, however they generally are applied to a pre-selected subspace, leaving out essential correlation.

Despite these advances, two central challenges remain. First, the identification of an appropriate active space becomes increasingly difficult as systems grow in size or complexity. For example, in extended $\pi$-conjugated systems such as polyacenes, the singlet--triplet gap decreases with system size while the separation between $\pi$ and $\sigma$ subspaces diminishes, and there is an increase multireference and diradical character within the $\pi$ subspace ~\cite{ST-delicate,ST-diradical,ST-reichman,yang2016nature,qu2009open}. As a result, the number of orbitals required to accurately describe the excitation grows, and so too does the cost of both selecting and solving the active space. Second, even when an appropriate active space can be identified, the computational cost of solving the corresponding many-body problem remains prohibitive for large systems. 

To recover the remaining correlation, recent developments have moved toward augmenting active-space treatments with density functional, perturbation, or adiabatic connection theories \cite{C8SC03569E,veis-pernal-adiabatic,ghosh2017generalized,li2014multiconfiguration,pernal2018electron,beran2021density,andersson1992second,angeli2001introduction}. These techniques belong to a broader class of approaches including dynamical downfolding and the constrained random phase approximation (cRPA) that incorporate external degrees of freedom by constructing effective Hamiltonians \cite{Aryasetiawan2009,SakumaWerner2013Dynamical,MiyakeEffevtiveBand,Romanova2023,Ma2021EmbeddingcRPA,DvorakRinke2019(ED),DvorakRinkeGolze2019ED,Sheng2022GFEmbedding,chang2024downfolding,chang2024renormalized,wagnerDMD,cDFT1,CDFT2}. While these methods effectively map the influence of the environment onto a reduced space, they still require the prior identification of a suitable active space and often introduce additional complexity, such as frequency-dependent interactions or self-energy terms that must be treated within the many-body solver.

These challenges motivate alternative formulations in which the computational bottleneck is not the size of the full Hamiltonian, but the size of the subsystems that must be solved. In this work, we adopt such an approach by expressing many-body observables as a cluster expansion over contributions from subsets of single-particle orbitals. In this formulation, the properties of the full system are reconstructed from a series of calculations on reduced-rank Hamiltonians, avoiding the need to identify a single optimal active space\cite{canestraight2026stochasticclusterexpansionelectronic}. This expansion provides a systematically improvable framework for capturing the correlation energy. This energy, which arises from electron-electron interactions, can be formally expressed as

\begin{equation}
    \label{eq: Totalenergy}
    E_{(0)} = E_{\rm MF} + \varepsilon_c = E_{\rm MF} + \sum_n \binom{N}{n} (\delta\varepsilon_{c})_n,
\end{equation}
where the combinatorial prefactor accounts for the number of distinct $n$-orbital groupings. Here, $(\delta\varepsilon_{c})_n$ denotes the contribution to the total correlation energy arising from all groupings of $n$ single-particle orbitals, and $E_0$ is the interacting ground-state energy. Cluster expansions of this form have been widely used across many areas of many-body physics\cite{PaesaniCluster1,PaesaniCluster2,Nesbet1968,Abraham_2021}.

For each term in the cluster expansion, the contribution to the correlation energy can be obtained by solving both the correlated and mean-field Hamiltonians and taking the difference in their total energies, thereby defining $(\delta\varepsilon_c)_n$ operationally. This implies that computation of the $N^{\rm th}$-order term formally requires solving the full many-body Hamiltonian, which is intractable for systems with hundreds to thousands of electrons. In practice, however, the cluster expansion may be truncated at lower order, providing an approximation to the total correlation energy while only requiring the solution of lower-rank Hamiltonians.

For systems in which higher-order couplings are important, such truncation may lead to systematic underestimation of the correlation energy. To address this, we exploit the locality of strong correlation by partitioning the single-particle orbital space into two components: a frontier chemical subspace (FCS), which is treated exactly and captures all orders of coupling, and a less-correlated rest space, for which the cluster expansion is truncated at low order\cite{canestraight2026stochasticclusterexpansionelectronic}. Expressed in terms of single-particle orbitals, this partitioned cluster expansion becomes:
\begin{equation}
\label{eq: correlationcluster}
        \varepsilon_c = \varepsilon_c^{\rm FCS} + \sum_\phi \delta\varepsilon_c^\phi + \sum_{\phi \phi'} \delta\varepsilon_c^{\phi \phi'} + \cdots,
\end{equation}
where the environment cluster terms are defined as
$\delta\varepsilon_c^{\phi} = \varepsilon_c^{\rm FCS + \phi} - \varepsilon_c^{\rm FCS}$ and
$\delta\varepsilon_c^{\phi \phi'} = \varepsilon_c^{\rm FCS + \phi + \phi'} - \varepsilon_c^{\rm FCS} - \delta\varepsilon_c^{\phi} - \delta\varepsilon_c^{\phi'}$.
These terms represent the change in the total correlation energy upon including one or two environment orbitals, $\phi$ and $\phi'$, in the active-space calculation alongside the FCS orbitals. In practice, these orbitals often come from mean-field calculations. In this work, we utilize an \emph{energy ordered} Kohn-Sham basis and choose this partitioning on the basis of proximity to the Fermi level. However, this method is compatible with more advanced methods for active space selection method such as CASSCF that can be applied to the mean-field single-particle basis. 

To obtain a many-body gap from the expression above, we take the difference between the total many-body energies computed from Eq.~\ref{eq: Totalenergy}:
\begin{equation}
\label{eq: many-bodygap}
E_{(\lambda)} - E_{(0)} = \Delta^{\rm FCS}_{(0,\lambda)} + \sum_\phi \delta \Delta^{\phi}_{(0,\lambda)} + \sum_{\phi,\phi'} \delta \Delta^{\phi, \phi'}_{(0,\lambda)} + \cdots.
\end{equation}
Here, $E_{(\lambda)}$ and $E_{(0)}$ denote two many-body states that can be reliably obtained using a solver such as exact diagonalization or DMRG. $\Delta^{\rm FCS}_{(0,\lambda)}$ is the gap computed within the FCS, and $\delta \Delta^{\phi}_{(0,\lambda)}$ represents the change in the gap upon inclusion of an additional orbital $\phi$ in the active space. Taking the difference eliminates the mean-field contribution, yielding an expression for the gap entirely in terms of FCS-based renormalizations. A full derivation of Eq.~\ref{eq: many-bodygap} is provided in the Supporting Information. Removing the dependence on mean-field total energy extends the applicability of the SCE to excited states that do not admit a well-defined mean-field reference, provided the excitation is predominantly captured within the FCS\footnote{A physically meaningful evaluation Eq.~\ref{eq: many-bodygap} requires that the gap is evaluated between states with (nearly) the same two eigenvectors for every stochastic sampling. When the excitation is not predominantly contained in the FCS, inclusion of stochastic orbitals $\phi$ and $\phi'$ may have a large impact on the eigenvectors of the many-body Hamiltonian, rendering the required eigenvector matching impossible.}.

This formulation already provides a substantial computational speed-up, as it replaces a single large many-body calculation with repeated gap evaluations in minimal active spaces, each of which has an exponentially smaller Hilbert space. Additional acceleration can be achieved by introducing a stochastic cluster expansion. In this approach, we include a sampled single-particle orbital that provides an unbiased representation of the rest space on average:
\begin{equation}
        |\zeta\rangle = \frac{1}{\sqrt{N_{\rm R}}} \sum_j^{N_{\rm R}} e^{i\theta_j} |\phi_j\rangle,
\end{equation}
where $\theta_j$ is randomly sampled from $[0,2\pi)$. For real-valued solvers, $\theta_j$ may be chosen such that the amplitudes remain real. In this case we take the $\pm \phi_j$ linear combination of states as has been employed in other methods\cite{Canestraight2024Efficient,apelian2024delocalization}.

One or more mutually orthogonal stochastic orbitals are then added to the FCS active space to estimate the average effect of environment orbitals. Rather than explicitly evaluating all environment contributions, these sampled contributions are weighted by the number of orbitals or orbital pairs they represent:
\begin{equation}
  \label{eq:stoch_gap}
        \langle \Delta_{(0,\lambda)} \rangle \approx \Delta_{(0,\lambda)}^{\rm FCS} + \left\langle N_{\rm R} \delta \Delta_{(0,\lambda)}^{\zeta} + \frac{N_{\rm R}(N_{\rm R}-1)}{2} \delta\Delta_{(0,\lambda)}^{\zeta \zeta'} \right\rangle_{N_{\zeta}}.
\end{equation}
Repeated sampling is required to converge the expectation value of the gap. This approach provides significant computational savings for redundant systems where many of the rest space orbitals interact with the gap in the same way, as well as for systems where only a subset of orbitals significantly renormalizes the gap but is not known \textit{a priori}.

In this work, we demonstrate the stochastic cluster expansion truncated at second order for molecular systems in which correlation is localized primarily to $\pi$ and $\pi^*$ orbitals nearest the Fermi level. The correlation energy obtained from the second order expansion is benchmarked against the DMRG result for all occupied $\pi$ and $\sigma$ orbitals, and it shows excellent agreement. The SCE method for many-body gaps is compatible with all many-body solvers. We utilize DMRG, yielding the ground state of the total spin $S=0$ and $S=1$ partitions, however, methods such as Configuration Interaction (CI) may be used instead\cite{DavidSherrill1999,Bartlett2007,Muller1934}. 
The SCE method has the advantage of treating both the inter-FCS orbital correlation and the FCS-environment correlation with the same solver. This, in principle, alleviates some of the method's dependence on the precise determination of the correlated subspace. The computational cost, however, is determined by the size of the selected correlated subspace, with each calculation utilizing an active space that includes the full FCS together with one or two stochastically sampled orbitals. A smaller orbital subspace implies an exponentially smaller Hilbert space for each calculation. We therefore benchmark the SCE method for excited states on a fully solvable model system to explore the impact of the number of occupied orbitals included in the FCS.

\begin{figure}
    \centering
    \includegraphics[width=\linewidth]{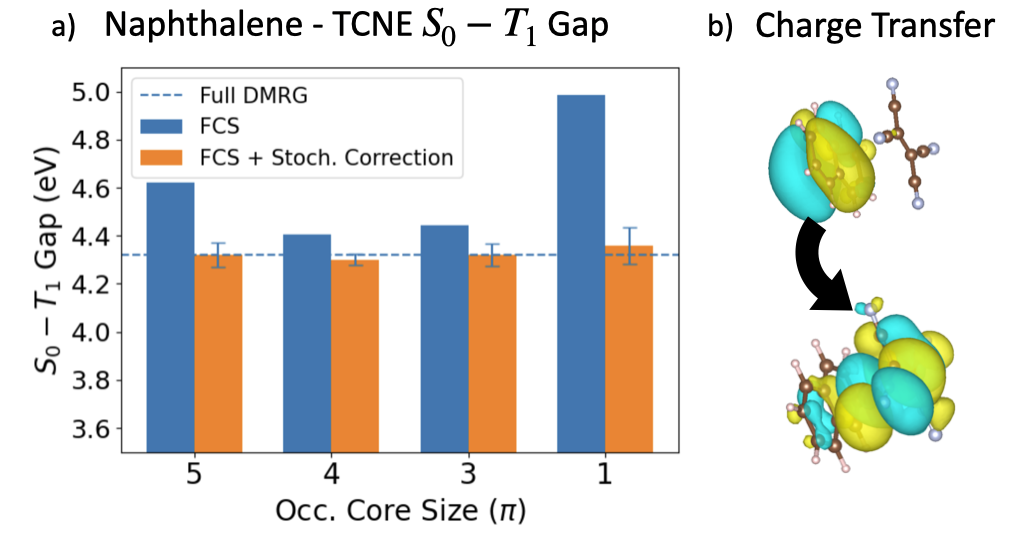}
    \caption{(a) The $S_0 - T_1$ gap computed for a range of FCS sizes in a Naphthalene--TCNE charge-transfer complex. Error bars denote the standard error of the mean for 25 samples. (b) The Naphthalene--TCNE complex, with the Naphthalene HOMO and TCNE LUMO shown.}
    \label{fig:fig1}
\end{figure}

For this benchmark, we consider a Naphthalene--Tetracyanoethylene (TCNE) charge-transfer complex\cite{dhar1967chemistry,C8SC03569E,SteinCT-complexes2009}, which we solve with DMRG for a basis of all $\sigma$, $\pi$, and $\pi^*$ orbitals. In this system, both molecules exhibit multiple overlapping $\pi$ and $\pi^*$ orbitals, and the triplet excitation leads to charge transfer from Naphthalene to TCNE. We define the FCS as consisting of the highest-energy occupied $\pi$ orbitals together with all $\pi^*$ orbitals from both molecules. In Fig.~\ref{fig:fig1}, we begin with five occupied $\pi$ orbitals in the FCS and progressively move them into the sampled rest space in order of decreasing energy. An FCS with one occupied $\pi$ orbital corresponds to the HOMO together with all $\pi^*$ orbitals. All orbitals excluded from the active space contribute a one-body Coulomb and exchange potential to the orbitals in the active space.

In Fig.~\ref{fig:fig1}, we observe that the singlet-triplet gap computed within the FCS alone does not match the full-system DMRG result (the blue horizontal line) for any of the simple orbital energy-based partitions demonstrated. When the stochastic correction is added to the FCS-only gap, however, the full DMRG result is recovered in all cases within the standard error of the mean (on average, $0.048 ~\rm eV$) for only 25 samples.

The success of the SCE correction across all FCS sizes indicates that the charge-transfer excitation is not strongly dependent on higher-order couplings. Since the stochastic cluster expansion truncated at second order recovers the full result across all partitionings, the dominant correlation effects can be captured through single-orbital and pairwise interactions. Pairwise interaction have similarly been found sufficient for ground-state correlation energy\cite{canestraight2026stochasticclusterexpansionelectronic}. For systems of this type, the SCE can therefore be applied using an FCS consisting only of the HOMO and the virtual space, resulting in a substantial reduction in the size of the largest Hamiltonian that must be solved. 

While the SCE recovers the singlet-triplet gap for all FCS-rest partitionings with as few as $25$ samples, the associated error bars differ, indicating that an optimal partitioning may exist that minimizes the number of samples required. This optimum partition may be system dependent. The efficiency of the method depends on the variance of the stochastic estimate for each partitioning. For example, although an occupied FCS size of one yields the smallest active space, it is also associated with the largest error bars. Multiple factors contribute to this convergence rate. The stochastic error is determined by a balance between the prefactors in Eq.~\ref{eq:stoch_gap} and the magnitude of the stochastic terms $\delta\Delta$. Due to the competing scaling of these prefactors and the $\delta\Delta$ terms, optimizing convergence is nontrivial. The prefactors decrease as the FCS is enlarged, while the magnitude of each $\delta\Delta$ term is reduced when sampling orbitals that weakly affect the gap.

To address this interplay, many stochastic approaches employ targeted sampling strategies that preferentially select orbitals expected to have the largest impact on the observable\cite{tubman2020modern,neuscamman2012optimizing,cleland2010communications,martin_2016}. In this work, we adopt a simple first-order prioritization scheme: we restrict the stochastic space to orbitals that, based on energetic and spatial proximity, are expected to most significantly influence the gap. This allows us to directly assess the impact of such selective sampling on both the observable and the convergence of its expectation value.

We benchmark this approach on two charge-transfer complexes: Benzene--TCNE and Naphthalene--TCNE. For both systems, we partition the single-particle orbitals based on energy and spatial localization on each molecule. This is done by performing Pipek-Mezey Wannierization on the system to obtain distinct sets of TCNE and acene orbitals (no hybridization). We then evaluate and diagonalize the Kohn Sham Hamiltonian (based on the full system density) acting on each molecular orbital block separately to obtain an approximate single-particle energy spectrum for each molecule. These spectra as well as further details of this orbital transformation are provided in the Supporting Information. In the resulting spectra from both complexes, TCNE $\pi$ orbitals lie energetically between the HOMO and HOMO$-1$ of the acene. 
When charge transfer occurs and an electron is promoted from the acene to the TCNE LUMO\cite{kuroda1967charge,blase2011charge,mei2019charge,kronik2012excitation}. We therefore define a \emph{targeted} scheme where the FCS consists of the acene $\pi$ orbitals and sampling is restricted to TCNE orbitals, as these are closer both in real-space and in energy to the LUMO orbital that receives the spin-flip electron. The remaining occupied acene $\sigma$ orbitals are treated as a mean-field potential. This is compared to a uniform sampling approach over the full rest space.

\begin{figure*}
    \centering
    \includegraphics[width=\textwidth]{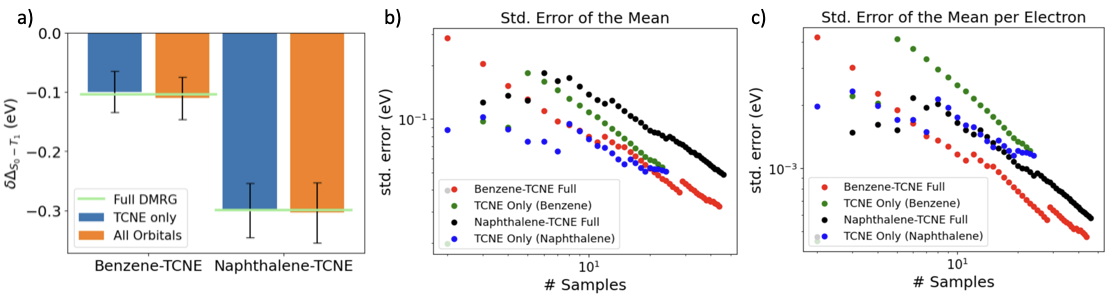}
    \caption{(a) Gap renormalization for Benzene--TCNE and Naphthalene--TCNE, compared with full DMRG. Error bars denote the standard error of the mean for 25 samples. (b) Standard error of the mean as a function of the number of stochastic samples. (c) Standard error of the mean per electron.}
    \label{fig:fig3}
\end{figure*}

In Fig.~\ref{fig:fig3}a, we plot the stochastic correction to the singlet--triplet gap for both systems. We find that restricting sampling to the TCNE orbitals yields the same gap renormalization as uniform sampling over the full rest space within the standard error of the mean for 25 samples. The horizontal line indicates the reference value obtained as the difference between full-system DMRG and FCS-only DMRG calculations.

Of greater interest is the magnitude of the error bars. No systematic advantage is observed when prioritized sampling is employed at a fixed sample number. Fig.~\ref{fig:fig3}b shows the stochastic error as a function of the number of samples. The standard error decreases with a slope consistent with $1/\sqrt{N_{\zeta}}$ (the rate of convergence found in QMC approaches \cite{martin_2016}), and uniform sampling provides no clear disadvantage relative to TCNE-only sampling. Fig.~\ref{fig:fig3}c shows the standard error per electron, where the uniform approach yields lower per-electron error. This indicates that self-averaging reduces the stochastic error on a per-electron basis. 

These results demonstrate that uniform sampling over the full rest space is as effective as chemically motivated targeted strategies for this class of systems. Convergence of the SCE does not benefit greatly from selective orbital prioritization, which is advantageous in practice: the method can be applied without \textit{a priori} knowledge of which orbitals most strongly influence the many-body gap. Based on the above results, we have established that the SCE for singlet-triplet gaps converges with a minimal FCS consisting of a single occupied orbital (the HOMO), and that this convergence is not strongly impacted by \textit{a priori} knowledge of which rest-space orbitals influence the gap. 

These properties enable the application of the method to significantly larger and more strongly correlated benchmark systems, such as the polyacenes, which are known to exhibit increasing diradical character with system size. Computation of polyacene singlet-triplet gaps is both a well-known challenge and a standard benchmark for many-body methods, including Auxiliary Field Quantum Monte Carlo (AFQMC), Density Matrix Renormalization Group (DMRG)\cite{MOLMPS,molmps_scalable} and related approaches (such as DMRG+Pair Density Functional Theory (PDFT)), as well as coupled cluster methods\cite{ST-delicate,ST-reichman,ST-diradical,hajgato2009benchmark,veis-pernal-adiabatic}. In these systems, the correlated singlet-triplet gap is primarily governed by the $\pi$-$\pi^*$ space, which lies symmetrically about the Fermi level. The number of $\pi$ orbitals grows linearly with the number of carbon rings, implying that the minimal active space required for conventional treatments grows exponentially with system size.

\begin{figure*}
    \centering
    \includegraphics[width=.75\textwidth]{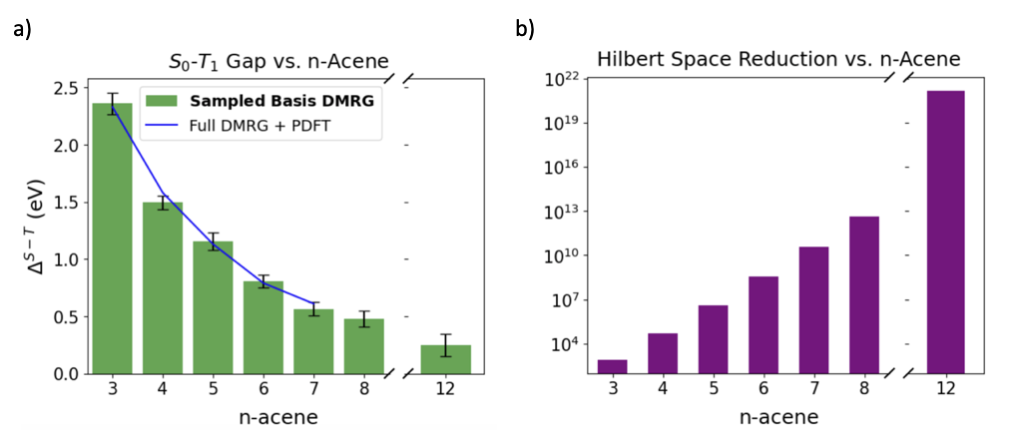}
    \caption{(a) Singlet--triplet gap as a function of $n$-acene length. Results are shown for the FCS-only calculation (HOMO and $\pi^*$), the SCE method, and DMRG+PDFT calculations\cite{C8SC03569E}. (b) Dimension of the full Hilbert space for the complete $\pi$ subspace compared to the largest active space used within the SCE method.}
    \label{fig:fig4}
\end{figure*}

For a range of polyacene sizes, we compute the singlet-triplet gap using a minimal FCS consisting of one occupied orbital (the HOMO) and all $\pi^*$ virtual orbitals. All lower-energy $\pi$ orbitals and $\sigma$ orbitals are sampled uniformly, consistent with the observation that unbiased sampling performs comparably to biased schemes. In Fig.~\ref{fig:fig4}a, we benchmark the SCE method against DMRG+PDFT results\cite{C8SC03569E} for systems up to heptacene. 

The SCE provides highly accurate results with substantially improved computational efficiency due to the reduced size of the active space. In Fig.~\ref{fig:fig4}b, we show the ratio of the dimension of the largest many-body Hilbert space arising in the SCE calculations to that obtained using the full $\pi$-$\pi^*$ space. For heptacene, our result shows excellent agreement with Ref.~\cite{C8SC03569E} in only 25 samples, while the dimension of the largest Hamiltonian that must be solved is reduced by $\sim 10$ orders of magnitude relative to the full $\pi$-$\pi^*$ active space. Although the SCE requires repeated evaluations to converge a statistical expectation value, the repeated solution of these low-rank Hamiltonians remains far less computationally demanding than a single solution in the full space.

As polyacene length increases, the energetic separation between the $\pi$ and $\sigma$ subspaces decreases, suggesting that $\sigma$ orbitals may play an increasing role in determining the singlet-triplet gap. The SCE method enables uniform treatment of all rest-space orbitals with minimal impact on convergence rate, allowing the gap to be computed robustly without \textit{a priori} knowledge of the relative importance of $\sigma$ orbitals. Furthermore, incorporation of these additional correlations does not require any increase in the size of the active space (FCS plus sampled orbitals).

Beyond computational efficiency, the success of the method provides insight into the nature of correlation in polyacenes. Although the HOMO--LUMO gap decreases and the number of $\pi$ orbitals increases with system size, the effective ``order'' of correlation, defined as the maximum orbital grouping size required for convergence of the cluster expansion within the SCE framework, remains constant. In particular, we recover the singlet-triplet gap using cluster contributions involving at most three orbitals (one in the FCS and two sampled). This suggests that the computational bottleneck in these systems arises from the growth of the orbital space rather than entanglement growth with the system size.

Finally, for sufficiently long polyacenes and related conjugated systems, the triplet excitation energy is known to decrease and can approach that of the singlet ground state\cite{jiang2008electronic}. In this regime, the ground state may acquire increasing open-shell character, and constructing a single-particle orbital basis from a closed-shell mean-field calculation may become suboptimal. Importantly, the present formulation places no restriction on the choice of single-particle orbital basis and remains applicable even in situations where singlet–triplet gaps become small or negative. Additionally, in the Supporting Information for the simplest case $\mathrm{O}_2$, the correct gap and triplet ground state can still be recovered even when starting from closed-shell orbitals.

To summarize, the stochastic cluster expansion (SCE) method provides a cost-effective and flexible framework for correlated electronic structure calculations. In this work, we have demonstrated its performance for singlet--triplet gaps in polyacenes, where it yields results comparable to DMRG and DMRG+PDFT with a substantially reduced-dimension Hamiltonian. Importantly, the method does not require chemically motivated partitioning of the active space or prior knowledge of which orbitals must be treated explicitly. Instead, all occupied orbitals may be treated on equal footing within the stochastic sampling framework without increasing the cost per sample.

A key practical advantage of the SCE framework is that its formulation does not restrict it to use with a single many-body solver. Although we have employed the MOLMPS DMRG\cite{MOLMPS,molmps_scalable} implementation to solve each sampled Hamiltonian, it has potential compatibility with a variety of other solvers, and exploring their use with the SCE provides a direction for future work. This flexibility allows the method to be adapted across different correlation regimes and system sizes.

While the present study focuses on vertical excitations, for which both states are evaluated in a common orbital basis and Hamiltonian, the formalism can be extended to more general settings. For adiabatic excitations, separate stochastic samplings would be required for the ground- and excited-state orbital bases, and the expansion must be carried out independently for each total energy through the correlation energy, since the gap formulation no longer benefits from cancellation of the mean-field reference. In this case, the stochastic errors in the two energies are statistically independent, and no covariance reduces the uncertainty in the gap, necessitating additional sampling to achieve comparable precision. Nonetheless, this methodology can still be applied to adiabatic excitations.

Further, the SCE formalism is, in principle, applicable to excited states beyond spin-defined manifolds. The singlet ($S_0$) and triplet ($T_1$) states considered here are uniquely defined as the lowest-energy eigenstates within their respective spin sectors, which avoids ambiguity in tracking eigenvectors across stochastic realizations. For more general excited states, consistent identification of a target eigenvector is required even as eigenvalues fluctuate between samples\cite{pineda2019excited,zhao2016efficient}. This challenge may be mitigated when the excitation can be approximated as a product state between the frontier chemical subspace and the ground state of the sampled environment, such that the environment renormalizes the excitation energy while leaving the eigenvector structure largely intact~\cite{CanestraightRenorm2025}.

This perspective highlights strong potential for systems in which excitations are localized to a subset of orbitals. Examples include defect centers and condensed-phase molecular systems, where excitation energies are dynamically renormalized by their surroundings but retain a predominantly local character\cite{Romanova2023}. In such cases, the SCE framework provides a natural route to treating localized excitations embedded within extended correlated environments. We view study of such systems as a direction for future work.

\section*{Acknowledgements}
  This material is based upon work supported by the U.S. Department of Energy, Office of Science, Office of Basic Energy Sciences, Computational and Theoretical Chemistry program under Award Number DE-SC0026045. The collaborative work on the implementation (AC, RM, LV, VV) was supported by Wellcome Leap as part of the Quantum for Bio Program. LV further acknowledges support from the Ministry of Education of the Czech Republic MSMT-CR, through the INTER-EXCELLENCE II, INTER-ACTION [LUAUS25286], and the e-INFRA CZ (ID:90254).

\bibliography{bib,bibnew}

@Article{Romanova2023,
author={Romanova, Mariya
and Weng, Guorong
and Apelian, Arsineh
and Vl{\v{c}}ek, Vojt{\v{e}}ch},
title={Dynamical downfolding for localized quantum states},
journal={npj Computational Materials},
year={2023},
month={Jul},
day={19},
volume={9},
number={1},
pages={126},
issn={2057-3960},
doi={10.1038/s41524-023-01078-5},
url={https://doi.org/10.1038/s41524-023-01078-5}
}

@article{GwenReducedScaling,
author = {Weng, Guorong and Romanova, Mariya and Apelian, Arsineh and Song, Hanbin and Vlček, Vojtěch},
title = {Reduced Scaling of Optimal Regional Orbital Localization via Sequential Exhaustion of the Single-Particle Space},
journal = {Journal of Chemical Theory and Computation},
volume = {18},
number = {8},
pages = {4960-4972},
year = {2022},
doi = {10.1021/acs.jctc.2c00315},
    note ={PMID: 35817013},

URL = { 
    
        https://doi.org/10.1021/acs.jctc.2c00315
    
    

},
eprint = { 
    
        https://doi.org/10.1021/acs.jctc.2c00315
    
    

}

}

@article{GwenEfficientTreatment,
    author = {Weng, Guorong and Vlček, Vojtěch},
    title = {Efficient treatment of molecular excitations in the liquid phase environment via stochastic many-body theory},
    journal = {The Journal of Chemical Physics},
    volume = {155},
    number = {5},
    pages = {054104},
    year = {2021},
    month = {08},
    issn = {0021-9606},
    doi = {10.1063/5.0058410},
    url = {https://doi.org/10.1063/5.0058410},
    eprint = {https://pubs.aip.org/aip/jcp/article-pdf/doi/10.1063/5.0058410/14111299/054104\_1\_online.pdf},
}

@article{PaesaniCluster1,
    author = {Medders, Gregory R. and Götz, Andreas W. and Morales, Miguel A. and Bajaj, Pushp and Paesani, Francesco},
    title = {On the representation of many-body interactions in water},
    journal = {The Journal of Chemical Physics},
    volume = {143},
    number = {10},
    pages = {104102},
    year = {2015},
    month = {09},
    issn = {0021-9606},
    doi = {10.1063/1.4930194},
    url = {https://doi.org/10.1063/1.4930194},
    eprint = {https://pubs.aip.org/aip/jcp/article-pdf/doi/10.1063/1.4930194/15502073/104102\_1\_online.pdf},
}

@article{PaesaniCluster2,
author = {Medders, Gregory R. and Babin, Volodymyr and Paesani, Francesco},
title = {A Critical Assessment of Two-Body and Three-Body Interactions in Water},
journal = {Journal of Chemical Theory and Computation},
volume = {9},
number = {2},
pages = {1103-1114},
year = {2013},
doi = {10.1021/ct300913g},
    note ={PMID: 26588754},

URL = { 
    
        https://doi.org/10.1021/ct300913g
    
    

},
eprint = { 
    
        https://doi.org/10.1021/ct300913g
    
    

}

}

@article{CanestraightRenorm2025,
    author = {Canestraight, Annabelle and Huang, Zhen and Lin, Lin and Vlcek, Vojtech},
    title = {Renormalization of states and quasiparticles in many-body downfolding},
    journal = {The Journal of Chemical Physics},
    volume = {163},
    number = {2},
    pages = {024114},
    year = {2025},
    month = {07},
    issn = {0021-9606},
    doi = {10.1063/5.0276233},
    url = {https://doi.org/10.1063/5.0276233},
    
}

@article{Canestraight2024Efficient,
author = {Canestraight, Annabelle and Lei, Xiaohe and Ibrahim, Khaled Z. and Vlček, Vojtěch},
title = {Efficient Quasiparticle Determination beyond the Diagonal Approximation via Random Compression},
journal = {Journal of Chemical Theory and Computation},
volume = {20},
number = {2},
pages = {551-557},
year = {2024},
doi = {10.1021/acs.jctc.3c01069},
    note ={PMID: 38175913},
URL = {   https://doi.org/10.1021/acs.jctc.3c01069
},}

@article{MOLMPS,
author = {Brabec, Jiri and Brandejs, Jan and Kowalski, Karol and Xantheas, Sotiris and Legeza, Ors and Veis, Libor},
title = {Massively parallel quantum chemical density matrix renormalization group method},
journal = {Journal of Computational Chemistry},
volume = {42},
number = {8},
pages = {534-544},
doi = {https://doi.org/10.1002/jcc.26476},
url = {https://onlinelibrary.wiley.com/doi/abs/10.1002/jcc.26476},
eprint = {https://onlinelibrary.wiley.com/doi/pdf/10.1002/jcc.26476},
year = {2021}
}

@article{OlivaresAmaya2015,
  title = {The ab-initio density matrix renormalization group in practice},
  volume = {142},
  ISSN = {1089-7690},
  url = {http://dx.doi.org/10.1063/1.4905329},
  DOI = {10.1063/1.4905329},
  number = {3},
  journal = {The Journal of Chemical Physics},
  publisher = {AIP Publishing},
  author = {Olivares-Amaya,  Roberto and Hu,  Weifeng and Nakatani,  Naoki and Sharma,  Sandeep and Yang,  Jun and Chan,  Garnet Kin-Lic},
  year = {2015},
  month = jan 
}

@article{Szalay2015,
  doi = {10.1002/qua.24898},
  url = {https://doi.org/10.1002/qua.24898},
  year = {2015},
  month = may,
  publisher = {Wiley},
  volume = {115},
  number = {19},
  pages = {1342--1391},
  author = {Szil{\'{a}}rd Szalay and Max Pfeffer and Valentin Murg and Gergely Barcza and Frank Verstraete and Reinhold Schneider and \"{O}rs Legeza},
  title = {Tensor product methods and entanglement optimization forab initioquantum chemistry},
  journal = {Int. J. Quant. Chem.}
}

@article{Legeza2003,
  doi = {10.1103/physrevb.68.195116},
  url = {https://doi.org/10.1103/physrevb.68.195116},
  year = {2003},
  month = nov,
  publisher = {American Physical Society ({APS})},
  volume = {68},
  number = {19},
  author = {\"{O}. Legeza and J. S{\'{o}}lyom},
  title = {Optimizing the density-matrix renormalization group method using quantum information entropy},
  journal = {Phys. Rev. B}
}

@Article{legeza_2003a,
  author = {{\"O}. Legeza and J. R\"oder and B. Hess},
  journal= {Phys. Rev. B},
  volume = {67},
  pages={125114},
  year=2003
}

@article{DMRG_White,
  title = {Density matrix formulation for quantum renormalization groups},
  author = {White, Steven R.},
  journal = {Phys. Rev. Lett.},
  volume = {69},
  issue = {19},
  pages = {2863--2866},
  numpages = {0},
  year = {1992},
  month = {Nov},
  publisher = {American Physical Society},
  doi = {10.1103/PhysRevLett.69.2863},
  url = {https://link.aps.org/doi/10.1103/PhysRevLett.69.2863}
}

@article{chan_review,
  author = {Chan, Garnet Kin-Lic and Sharma, Sandeep},
  title = {The Density Matrix Renormalization Group in Quantum Chemistry},
  journal = {Annu. Rev. Phys. Chem.},
  volume = {62},
  number = {1},
  pages = {465-481},
  year = {2011},
}

@article{reiher_perspective,
  doi = {10.1063/1.5129672},
  url = {https://doi.org/10.1063/1.5129672},
  year = {2020},
  month = jan,
  publisher = {{AIP} Publishing},
  volume = {152},
  number = {4},
  pages = {040903},
  author = {Alberto Baiardi and Markus Reiher},
  title = {The density matrix renormalization group in chemistry and molecular physics: Recent developments and new challenges},
  journal = {J. Chem. Phys.}
}

@article{Bartlett2007,
  title = {Coupled-cluster theory in quantum chemistry},
  volume = {79},
  ISSN = {1539-0756},
  url = {http://dx.doi.org/10.1103/RevModPhys.79.291},
  DOI = {10.1103/revmodphys.79.291},
  number = {1},
  journal = {Reviews of Modern Physics},
  publisher = {American Physical Society (APS)},
  author = {Bartlett,  Rodney J. and Musiał,  Monika},
  year = {2007},
  month = feb,
  pages = {291–352}
}

@inbook{DavidSherrill1999,
  title = {The Configuration Interaction Method: Advances in Highly Correlated Approaches},
  ISSN = {0065-3276},
  url = {http://dx.doi.org/10.1016/S0065-3276(08)60532-8},
  DOI = {10.1016/s0065-3276(08)60532-8},
  booktitle = {Advances in Quantum Chemistry},
  publisher = {Elsevier},
  author = {David Sherrill,  C. and Schaefer,  Henry F.},
  year = {1999},
  pages = {143–269}
}

@article{Muller1934,
  title = {Note on an Approximation Treatment for Many-Electron Systems},
  volume = {46},
  ISSN = {0031-899X},
  url = {http://dx.doi.org/10.1103/PhysRev.46.618},
  DOI = {10.1103/physrev.46.618},
  number = {7},
  journal = {Physical Review},
  publisher = {American Physical Society (APS)},
  author = {Møller,  Chr. and Plesset,  M. S.},
  year = {1934},
  month = oct,
  pages = {618–622}
}

@article{Abraham_2021,
   title={Cluster many-body expansion: A many-body expansion of the electron correlation energy about a cluster mean field reference},
   volume={155},
   ISSN={1089-7690},
   url={http://dx.doi.org/10.1063/5.0057752},
   DOI={10.1063/5.0057752},
   number={5},
   journal={The Journal of Chemical Physics},
   publisher={AIP Publishing},
   author={Abraham, Vibin and Mayhall, Nicholas J.},
   year={2021},
   month=aug }

@article{Nesbet1968,
  title = {Atomic Bethe-Goldstone Equations. III. Correlation Energies of Ground States of Be, B, C, N, O, F, and Ne},
  author = {Nesbet, R. K.},
  journal = {Phys. Rev.},
  volume = {175},
  issue = {1},
  pages = {2--9},
  numpages = {0},
  year = {1968},
  month = {Nov},
  publisher = {American Physical Society},
  doi = {10.1103/PhysRev.175.2},
  url = {https://link.aps.org/doi/10.1103/PhysRev.175.2}
}

@article{Aryasetiawan2009,
  title = {Downfolded Self-Energy of Many-Electron Systems},
  author = {Aryasetiawan, F. and Tomczak, J. M. and Miyake, T. and Sakuma, R.},
  journal = {Phys. Rev. Lett.},
  volume = {102},
  issue = {17},
  pages = {176402},
  numpages = {4},
  year = {2009},
  month = {Apr},
  publisher = {American Physical Society},
  doi = {10.1103/PhysRevLett.102.176402},
  url = {https://link.aps.org/doi/10.1103/PhysRevLett.102.176402}
}

@article{MiyakeEffevtiveBand,
  title = {Ab initio procedure for constructing effective models of correlated materials with entangled band structure},
  author = {Miyake, Takashi and Aryasetiawan, Ferdi and Imada, Masatoshi},
  journal = {Phys. Rev. B},
  volume = {80},
  issue = {15},
  pages = {155134},
  numpages = {6},
  year = {2009},
  month = {Oct},
  publisher = {American Physical Society},
  doi = {10.1103/PhysRevB.80.155134},
  url = {https://link.aps.org/doi/10.1103/PhysRevB.80.155134}
}

@article{SakumaWerner2013Dynamical,
  title = {Electronic structure of SrVO${}_{3}$ within $GW$+DMFT},
  author = {Sakuma, R. and Werner, Ph. and Aryasetiawan, F.},
  journal = {Phys. Rev. B},
  volume = {88},
  issue = {23},
  pages = {235110},
  numpages = {8},
  year = {2013},
  month = {Dec},
  publisher = {American Physical Society},
  doi = {10.1103/PhysRevB.88.235110},
  url = {https://link.aps.org/doi/10.1103/PhysRevB.88.235110}
}

@article{AFQMCEskridge2019,
author = {Eskridge, Brandon and Krakauer, Henry and Zhang, Shiwei},
title = {Local Embedding and Effective Downfolding in the Auxiliary-Field Quantum Monte Carlo Method},
journal = {Journal of Chemical Theory and Computation},
volume = {15},
number = {7},
pages = {3949-3959},
year = {2019},
doi = {10.1021/acs.jctc.8b01244},
    note ={PMID: 31244125},
URL = {  https://doi.org/10.1021/acs.jctc.8b01244},
eprint = { https://doi.org/10.1021/acs.jctc.8b01244}
}

@article{DvorakRinkeGolze2019ED,
  title = {Quantum embedding theory in the screened Coulomb interaction: Combining configuration interaction with $GW/\mathrm{BSE}$},
  author = {Dvorak, Marc and Golze, Dorothea and Rinke, Patrick},
  journal = {Phys. Rev. Mater.},
  volume = {3},
  issue = {7},
  pages = {070801},
  numpages = {6},
  year = {2019},
  month = {Jul},
  publisher = {American Physical Society},
  doi = {10.1103/PhysRevMaterials.3.070801},
  url = {https://link.aps.org/doi/10.1103/PhysRevMaterials.3.070801}
}

@article{Muchler2022Static,
  title = {Quantum embedding methods for correlated excited states of point defects: Case studies and challenges},
  author = {Muechler, Lukas and Badrtdinov, Danis I. and Hampel, Alexander and Cano, Jennifer and R\"osner, Malte and Dreyer, Cyrus E.},
  journal = {Phys. Rev. B},
  volume = {105},
  issue = {23},
  pages = {235104},
  numpages = {19},
  year = {2022},
  month = {Jun},
  publisher = {American Physical Society},
  doi = {10.1103/PhysRevB.105.235104},
  url = {https://link.aps.org/doi/10.1103/PhysRevB.105.235104}
}

@misc{shee2024staticquantumembeddingscheme,
      title={A static quantum embedding scheme based on coupled cluster theory}, 
      author={Avijit Shee and Fabian M. Faulstich and Birgitta Whaley and Lin Lin and Martin Head-Gordon},
      year={2024},
      eprint={2404.09078},
      archivePrefix={arXiv},
      primaryClass={physics.chem-ph},
      url={https://arxiv.org/abs/2404.09078}, 
}

@misc{NormSquishActiveSapce,
      title={Self-consistent Quantum Iteratively Sparsified Hamiltonian method (SQuISH): A new algorithm for efficient Hamiltonian simulation and compression}, 
      author={Diana B. Chamaki and Stuart Hadfield and Katherine Klymko and Bryan O'Gorman and Norm M. Tubman},
      year={2022},
      eprint={2211.16522},
      archivePrefix={arXiv},
      primaryClass={quant-ph},
      url={https://arxiv.org/abs/2211.16522}, 
}

@ARTICLE{He2020-raFOEmbedding,
  title    = "A zeroth-order active-space frozen-orbital embedding scheme for
              multireference calculations",
  author   = "He, Nan and Evangelista, Francesco A",
  journal  = "The Journal of Chemical Physics",
  volume   =  152,
  number   =  9,
  pages    = "094107",
  month    =  mar,
  year     =  2020
}

@ARTICLE{Ma2021EmbeddingcRPA,
  title     = "Quantum Embedding Theory for Strongly Correlated States in
               Materials",
  author    = "Ma, He and Sheng, Nan and Govoni, Marco and Galli, Giulia",
  journal   = "J. Chem. Theory Comput.",
  publisher = "American Chemical Society",
  volume    =  17,
  number    =  4,
  pages     = "2116--2125",
  month     =  apr,
  year      =  2021
}

@ARTICLE{Sheng2022GFEmbedding,
  title     = "Green's Function Formulation of Quantum Defect Embedding Theory",
  author    = "Sheng, Nan and Vorwerk, Christian and Govoni, Marco and Galli,
               Giulia",
  journal   = "J. Chem. Theory Comput.",
  publisher = "American Chemical Society",
  volume    =  18,
  number    =  6,
  pages     = "3512--3522",
  month     =  jun,
  year      =  2022
}

@article{veis-pernal-adiabatic,
author = {Drwal, Daria and Beran, Pavel and Hapka, Michał and Modrzejewski, Marcin and Sokół, Adam and Veis, Libor and Pernal, Katarzyna},
title = {Efficient Adiabatic Connection Approach for Strongly Correlated Systems: Application to Singlet–Triplet Gaps of Biradicals},
journal = {The Journal of Physical Chemistry Letters},
volume = {13},
number = {20},
pages = {4570-4578},
year = {2022},
doi = {10.1021/acs.jpclett.2c00993},
    note ={PMID: 35580342},

URL = { 
    
        https://doi.org/10.1021/acs.jpclett.2c00993
    
    

},
eprint = { 
    
        https://doi.org/10.1021/acs.jpclett.2c00993
    
    

}

}

@article{SteinCT-complexes2009,
author = {Stein, Tamar and Kronik, Leeor and Baer, Roi},
title = {Reliable Prediction of Charge Transfer Excitations in Molecular Complexes Using Time-Dependent Density Functional Theory},
journal = {Journal of the American Chemical Society},
volume = {131},
number = {8},
pages = {2818-2820},
year = {2009},
doi = {10.1021/ja8087482},
    note ={PMID: 19239266},

URL = { 
    
        https://doi.org/10.1021/ja8087482
},
eprint = {  https://doi.org/10.1021/ja8087482
}}

@Article{C8SC03569E,
author ="Sharma, Prachi and Bernales, Varinia and Knecht, Stefan and Truhlar, Donald G. and Gagliardi, Laura",
title  ="Density matrix renormalization group pair-density functional theory (DMRG-PDFT): singlet–triplet gaps in polyacenes and polyacetylenes",
journal  ="Chem. Sci.",
year  ="2019",
volume  ="10",
issue  ="6",
pages  ="1716-1723",
publisher  ="The Royal Society of Chemistry",
doi  ="10.1039/C8SC03569E",
url  ="http://dx.doi.org/10.1039/C8SC03569E"}

@article{dhar1967chemistry,
  title={The Chemistry of tetracyanoethylene},
  author={Dhar, Durga Nath},
  journal={Chemical Reviews},
  volume={67},
  number={6},
  pages={611--622},
  year={1967},
  publisher={ACS Publications}
}

@article{ST-reichman
, author = {Shee, James and Arthur, Evan J. and Zhang, Shiwei and Reichman, David R. and Friesner, Richard A.},

title = {Singlet-Triplet Energy Gaps of Organic Biradicals and Polyacenes with Auxiliary-Field Quantum Monte Carlo},
journal = {Journal of Chemical Theory and Computation},
volume = {15},
number = {9},
pages = {4924-4932},
year = {2019},
doi = {10.1021/acs.jctc.9b00534},
    note ={PMID: 31381324},

URL = { https://doi.org/10.1021/acs.jctc.9b00534},}

@article{ST-diradical,
    author = {Hachmann, Johannes and Dorando, Jonathan J. and Avilés, Michael and Chan, Garnet Kin-Lic},
    title = {The radical character of the acenes: A density matrix renormalization group study},
    journal = {The Journal of Chemical Physics},
    volume = {127},
    number = {13},
    pages = {134309},
    year = {2007},
    month = {10},
    issn = {0021-9606},
    doi = {10.1063/1.2768362},
    url = {https://doi.org/10.1063/1.2768362},
   
}

@Article{ST-delicate,
author ="Ibeji, Collins U. and Ghosh, Debashree",
title  ="Singlet–triplet gaps in polyacenes: a delicate balance between dynamic and static correlations investigated by spin–flip methods",
journal  ="Phys. Chem. Chem. Phys.",
year  ="2015",
volume  ="17",
issue  ="15",
pages  ="9849-9856",
publisher  ="The Royal Society of Chemistry",
doi  ="10.1039/C5CP00214A",
url  ="http://dx.doi.org/10.1039/C5CP00214A"}

@article{WeirdAlgorithmiq,
author = {Kolodzeiski, Elena and Stein, Christopher J.},
title = {Automated, Consistent, and Even-Handed Selection of Active Orbital Spaces for Quantum Embedding},
journal = {Journal of Chemical Theory and Computation},
volume = {19},
number = {19},
pages = {6643-6655},
year = {2023},
doi = {10.1021/acs.jctc.3c00653},
    note ={PMID: 37775093},

URL = {        https://doi.org/10.1021/acs.jctc.3c00653
},
eprint = { 
    https://doi.org/10.1021/acs.jctc.3c00653
}

}

@article{canestraight2026stochasticclusterexpansionelectronic,
author = {Canestraight, Annabelle and Dominic III, Anthony J.and Montoya-Castillo, Andr{\'e}s and Veis, Libor and Vlcek, Vojtech},
title = {A Stochastic Cluster Expansion for Electronic Correlation in Large Systems},
journal = {The Journal of Physical Chemistry Letters},
volume = {0},
number = {0},
pages = {null},
year = {0},
doi = {10.1021/acs.jpclett.6c00563},

URL = { 
    
        https://doi.org/10.1021/acs.jpclett.6c00563
    
    

},
eprint = { 
    
        https://doi.org/10.1021/acs.jpclett.6c00563
    
    

}

}

@article{eriksen2020shape,
  title={The shape of full configuration interaction to come},
  author={Eriksen, Janus J},
  journal={The Journal of Physical Chemistry Letters},
  volume={12},
  number={1},
  pages={418--432},
  year={2020},
  publisher={ACS Publications}
}

@article{malmqvist1989casscf,
  title={The CASSCF state interaction method},
  author={Malmqvist, Per-{\AA}ke and Roos, Bj{\"o}rn O},
  journal={Chemical physics letters},
  volume={155},
  number={2},
  pages={189--194},
  year={1989},
  publisher={Elsevier}
}

@article{olsen2011casscf,
  title={The CASSCF method: A perspective and commentary},
  author={Olsen, Jeppe},
  journal={International Journal of Quantum Chemistry},
  volume={111},
  number={13},
  pages={3267--3272},
  year={2011},
  publisher={Wiley Online Library}
}

@article{helmich2019benchmarks,
  title={Benchmarks for electronically excited states with CASSCF methods},
  author={Helmich-Paris, Benjamin},
  journal={Journal of Chemical Theory and Computation},
  volume={15},
  number={7},
  pages={4170--4179},
  year={2019},
  publisher={ACS Publications}}

@article{jiang2008electronic,
  title={Electronic ground state of higher acenes},
  author={Jiang, De-en and Dai, Sheng},
  journal={The Journal of Physical Chemistry A},
  volume={112},
  number={2},
  pages={332--335},
  year={2008},
  publisher={ACS Publications}
}

@article{yang2016nature,
  title={Nature of ground and electronic excited states of higher acenes},
  author={Yang, Yang and Davidson, Ernest R and Yang, Weitao},
  journal={Proceedings of the National Academy of Sciences},
  volume={113},
  number={35},
  pages={E5098--E5107},
  year={2016},
  publisher={National Academy of Sciences}
}

@article{hajgato2009benchmark,
  title={A benchmark theoretical study of the electronic ground state and of the singlet-triplet split of benzene and linear acenes},
  author={Hajgat{\'o}, Balazs and Szieberth, D and Geerlings, Paul and De Proft, Frank and Deleuze, MS},
  journal={The Journal of chemical physics},
  volume={131},
  number={22},
  year={2009},
  publisher={AIP Publishing}
}

@article{qu2009open,
  title={Open-shell ground state of polyacenes: a valence bond study},
  author={Qu, Zexing and Zhang, Dawei and Liu, Chungen and Jiang, Yuansheng},
  journal={The Journal of Physical Chemistry A},
  volume={113},
  number={27},
  pages={7909--7914},
  year={2009},
  publisher={ACS Publications}
}

@article{sayfutyarova2017automated,
  title={Automated construction of molecular active spaces from atomic valence orbitals},
  author={Sayfutyarova, Elvira R and Sun, Qiming and Chan, Garnet Kin-Lic and Knizia, Gerald},
  journal={Journal of chemical theory and computation},
  volume={13},
  number={9},
  pages={4063--4078},
  year={2017},
  publisher={ACS Publications}
}

@article{bao2019automatic,
  title={Automatic active space selection for calculating electronic excitation energies based on high-spin unrestricted Hartree--Fock orbitals},
  author={Bao, Jie J and Truhlar, Donald G},
  journal={Journal of Chemical Theory and Computation},
  volume={15},
  number={10},
  pages={5308--5318},
  year={2019},
  publisher={ACS Publications}
}

@article{chang2024downfolding,
  title={Downfolding from ab initio to interacting model Hamiltonians: comprehensive analysis and benchmarking of the DFT+ cRPA approach},
  author={Chang, Yueqing and van Loon, Erik GCP and Eskridge, Brandon and Busemeyer, Brian and Morales, Miguel A and Dreyer, Cyrus E and Millis, Andrew J and Zhang, Shiwei and Wehling, Tim O and Wagner, Lucas K and others},
  journal={npj Computational Materials},
  volume={10},
  number={1},
  pages={129},
  year={2024},
  publisher={Nature Publishing Group UK London}
}

@article{chang2024renormalized,
  title={Renormalized density matrix downfolding: A rigorous framework in learning emergent models from ab initio many-body calculations},
  author={Chang, Yueqing and Joshi, Sonali and Wagner, Lucas K},
  journal={Physical Review B},
  volume={110},
  number={19},
  pages={195103},
  year={2024},
  publisher={APS}
}

@article{cDFT1,
  title = {Ground States of Constrained Systems: Application to Cerium Impurities},
  author = {Dederichs, P. H. and Bl\"ugel, S. and Zeller, R. and Akai, H.},
  journal = {Phys. Rev. Lett.},
  volume = {53},
  issue = {26},
  pages = {2512--2515},
  numpages = {0},
  year = {1984},
  month = {Dec},
  publisher = {American Physical Society},
  doi = {10.1103/PhysRevLett.53.2512},
  url = {https://link.aps.org/doi/10.1103/PhysRevLett.53.2512}
}

@article{CDFT2,
  title = {Density-functional calculation of the parameters in the Anderson model: Application to Mn in CdTe},
  author = {Gunnarsson, O. and Andersen, O. K. and Jepsen, O. and Zaanen, J.},
  journal = {Phys. Rev. B},
  volume = {39},
  issue = {3},
  pages = {1708--1722},
  numpages = {0},
  year = {1989},
  month = {Jan},
  publisher = {American Physical Society},
  doi = {10.1103/PhysRevB.39.1708},
  url = {https://link.aps.org/doi/10.1103/PhysRevB.39.1708}
}

@ARTICLE{wagnerDMD,
    
AUTHOR={Zheng, Huihuo  and Changlani, Hitesh J.  and Williams, Kiel T.  and Busemeyer, Brian  and Wagner, Lucas K. },
           
TITLE={From Real Materials to Model Hamiltonians With Density Matrix Downfolding},
          
JOURNAL={Frontiers in Physics},
          
VOLUME={Volume 6 - 2018},
  
YEAR={2018},
  
URL={https://www.frontiersin.org/journals/physics/articles/10.3389/fphy.2018.00043},
  
DOI={10.3389/fphy.2018.00043},
  
ISSN={2296-424X}}

@article{kasha1950characterization,
  title={Characterization of electronic transitions in complex molecules},
  author={Kasha, Michael},
  journal={Discussions of the Faraday society},
  volume={9},
  pages={14--19},
  year={1950},
  publisher={Royal Society of Chemistry}
}

@article{kasha1960paths,
  title={Paths of molecular excitation},
  author={Kasha, Michael},
  journal={Radiation research supplement},
  volume={2},
  pages={243--275},
  year={1960},
  publisher={JSTOR}
}

@book{Lakowicz2006,
  title     = {Principles of Fluorescence Spectroscopy},
  author    = {Lakowicz, Joseph R.},
  year      = {2006},
  publisher = {Springer},
  address   = {Singapore},
  edition   = {3rd},
  doi       = {10.1007/978-0-387-46312-4},
  isbn      = {978-0-387-31278-1}
}

@article{Zimmerman2010,
  title={Singlet fission in pentacene through multi-exciton quantum states},
  author={Zimmerman, Paul M. and Zhang, Zhiyong and Musgrave, Charles B.},
  journal={Nature Chemistry},
  volume={2},
  number={8},
  pages={648--652},
  year={2010},
  publisher={Nature Publishing Group}
}

@article{Uoyama2012,
  title={Highly efficient organic light-emitting diodes from delayed fluorescence},
  author={Uoyama, Hiroki and Goushi, Kenichi and Shizu, Katsuyuki and Nomura, Hiroko and Adachi, Chihaya},
  journal={Nature},
  volume={492},
  number={7428},
  pages={234--238},
  year={2012},
  publisher={Nature Publishing Group}
}

@article{Ma2010Modeling,
author = {Ma, Yuchen and Rohlfing, Michael and Molteni, Carla},
title = {Modeling the Excited States of Biological Chromophores within Many-Body Green's Function Theory},
journal = {Journal of Chemical Theory and Computation},
volume = {6},
number = {1},
pages = {257-265},
year = {2010},
doi = {10.1021/ct900528h},
    note ={PMID: 26614336},

URL = { 
    
        https://doi.org/10.1021/ct900528h
    
    

},
eprint = { 
    
        https://doi.org/10.1021/ct900528h
    
    

}

}

@article{apelian2024delocalization,
  title={Delocalization of Quasiparticle Moir{\'e} States in Twisted Bilayer hBN},
  author={Apelian, Arsineh and Canestraight, Annabelle and Liu, Songyuan and Vlcek, Vojtech},
  journal={Nano Letters},
  volume={24},
  number={38},
  pages={11882--11888},
  year={2024},
  publisher={ACS Publications}
}

@article{beran2021density,
  title={Density matrix renormalization group with dynamical correlation via adiabatic connection},
  author={Beran, Pavel and Matousek, Mikulas and Hapka, Micha{\l} and Pernal, Katarzyna and Veis, Libor},
  journal={Journal of Chemical Theory and Computation},
  volume={17},
  number={12},
  pages={7575--7585},
  year={2021},
  publisher={ACS Publications}
}

@article{pernal2018electron,
  title={Electron correlation from the adiabatic connection for multireference wave functions},
  author={Pernal, Katarzyna},
  journal={Physical review letters},
  volume={120},
  number={1},
  pages={013001},
  year={2018},
  publisher={APS}
}

@article{ghosh2017generalized,
  title={Generalized-active-space pair-density functional theory: an efficient method to study large, strongly correlated, conjugated systems},
  author={Ghosh, Soumen and Cramer, Christopher J and Truhlar, Donald G and Gagliardi, Laura},
  journal={Chemical science},
  volume={8},
  number={4},
  pages={2741--2750},
  year={2017},
  publisher={Royal Society of Chemistry}
}

@article{li2014multiconfiguration,
  title={Multiconfiguration pair-density functional theory},
  author={Li Manni, Giovanni and Carlson, Rebecca K and Luo, Sijie and Ma, Dongxia and Olsen, Jeppe and Truhlar, Donald G and Gagliardi, Laura},
  journal={Journal of chemical theory and computation},
  volume={10},
  number={9},
  pages={3669--3680},
  year={2014},
  publisher={ACS Publications}
}

@article{tubman2020modern,
  title={Modern approaches to exact diagonalization and selected configuration interaction with the adaptive sampling CI method},
  author={Tubman, Norm M and Freeman, C Daniel and Levine, Daniel S and Hait, Diptarka and Head-Gordon, Martin and Whaley, K Birgitta},
  journal={Journal of chemical theory and computation},
  volume={16},
  number={4},
  pages={2139--2159},
  year={2020},
  publisher={ACS Publications}
}

@article{cleland2010communications,
  title={Communications: Survival of the fittest: Accelerating convergence in full configuration-interaction quantum Monte Carlo},
  author={Cleland, Deidre and Booth, George H and Alavi, Ali},
  journal={The Journal of chemical physics},
  volume={132},
  number={4},
  year={2010},
  publisher={AIP Publishing}
}

@article{neuscamman2012optimizing,
  title={Optimizing large parameter sets in variational quantum Monte Carlo},
  author={Neuscamman, Eric and Umrigar, CJ and Chan, Garnet Kin-Lic},
  journal={Physical Review B—Condensed Matter and Materials Physics},
  volume={85},
  number={4},
  pages={045103},
  year={2012},
  publisher={APS}
}

@book{martin_2016, place={Cambridge}, title={Interacting Electrons: Theory and Computational Approaches}, DOI={10.1017/CBO9781139050807}, publisher={Cambridge University Press}, author={Martin, Richard M. and Reining, Lucia and Ceperley, David M.}, year={2016}}

@article{kronik2012excitation,
  title={Excitation gaps of finite-sized systems from optimally tuned range-separated hybrid functionals},
  author={Kronik, Leeor and Stein, Tamar and Refaely-Abramson, Sivan and Baer, Roi},
  journal={Journal of Chemical Theory and Computation},
  volume={8},
  number={5},
  pages={1515--1531},
  year={2012},
  publisher={ACS Publications}
}

@article{blase2011charge,
  title={Charge-transfer excitations in molecular donor-acceptor complexes within the many-body Bethe-Salpeter approach},
  author={Blase, Xavier and Attaccalite, Claudio},
  journal={Applied Physics Letters},
  volume={99},
  number={17},
  year={2011},
  publisher={AIP Publishing}
}

@article{mei2019charge,
  title={Charge transfer excitation energies from ground state density functional theory calculations},
  author={Mei, Yuncai and Yang, Weitao},
  journal={The Journal of Chemical Physics},
  volume={150},
  number={14},
  year={2019},
  publisher={AIP Publishing}
}

@article{kuroda1967charge,
  title={Charge-transfer interaction in tetracyanoethylene complexes of pyrene and naphthalene},
  author={Kuroda, Haruo and Amano, Takako and Ikemoto, Isao and Akamatu, Hideo},
  journal={Journal of the American Chemical Society},
  volume={89},
  number={24},
  pages={6056--6063},
  year={1967},
  publisher={ACS Publications}
}

@article{zhao2016efficient,
  title={An efficient variational principle for the direct optimization of excited states},
  author={Zhao, Luning and Neuscamman, Eric},
  journal={Journal of chemical theory and computation},
  volume={12},
  number={8},
  pages={3436--3440},
  year={2016},
  publisher={ACS Publications}
}

@article{pineda2019excited,
  title={Excited state specific multi-Slater Jastrow wave functions},
  author={Pineda Flores, Sergio D and Neuscamman, Eric},
  journal={The Journal of Physical Chemistry A},
  volume={123},
  number={8},
  pages={1487--1497},
  year={2019},
  publisher={ACS Publications}
}

@article{andersson1992second,
  title={Second-order perturbation theory with a complete active space self-consistent field reference function},
  author={Andersson, Kerstin and Malmqvist, Per-{\AA}ke and Roos, Bj{\"o}rn O},
  journal={The Journal of chemical physics},
  volume={96},
  number={2},
  pages={1218--1226},
  year={1992},
  publisher={American Institute of Physics}
}

@article{angeli2001introduction,
  title={Introduction of n-electron valence states for multireference perturbation theory},
  author={Angeli, Celestino and Cimiraglia, Renzo and Evangelisti, Stefano and Leininger, Thierry and Malrieu, J-P},
  journal={The Journal of Chemical Physics},
  volume={114},
  number={23},
  pages={10252--10264},
  year={2001},
  publisher={American Institute of Physics}
}

@misc{molmps_scalable,
  author       = {{molmps developers}},
  title        = {Scalable},
  year         = {2026},
  howpublished = {\url{https://gitlab.com/molmps/scalable}},
  note         = {GitLab repository, accessed April 30, 2026}
}
\end{document}